\documentclass[sigconf,authorversion]{acmart}

\setcopyright{none}
\renewcommand\footnotetextcopyrightpermission[1]{}

\usepackage{array}

\begin{document}

\title{Digital Mentor: towards a conversational bot to identify hypotheses for software startups}

\author{Jorge Melegati}
\email{jorge.melegati@unibz.it}
\orcid{0000-0003-1303-4173}
\affiliation{%
  \institution{Free University of Bozen-Bolzano}
  \streetaddress{Piazza Domenicani 3}
  \city{Bolzano}
  \country{Italy}
  \postcode{39100}
}

\author{Xiaofeng Wang}
\email{xiaofeng.wang@unibz.it}
\orcid{0000-0001-8424-419X}
\affiliation{%
	\institution{Free University of Bozen-Bolzano}
	\streetaddress{Piazza Domenicani 3}
	\city{Bolzano}
	\country{Italy}
	\postcode{39100}
}


\begin{abstract}
Software startups develop innovative, software-intensive product and services. This context leads to uncertainty regarding the software they are building. Experimentation, a process of testing hypotheses about the product, helps these companies to reduce uncertainty through different evidence-based approaches. The first step in experimentation is to identify the hypotheses to be tested. HyMap is a technique where a facilitator helps a software startup founder to draw a cognitive map representing her understanding of the context and, based on that, create hypotheses about the software to be built. In this paper, we present the \textit{Digital Mentor}, an working-in-progress conversational bot to help creating a HyMap without the need of a human facilitator. We report the proposed solution consisting of a web application with the backend of a natural language understanding system, the current state of development, the challenges we faced so far and the next steps we plan to move forward. 
\end{abstract}




\pagestyle{plain}
\maketitle

\section{Introduction}
\label{sec:introduction}

Software startups are organizations searching for sustainable and scalable business model for innovative software-intensive products or services~\cite{Unterkalmsteiner2016}. The uncertainty associated with developing an innovative product is one of the reasons to a high rate of failure among these companies. In this regard, one of the key reasons for failure is not building a solution that potential customers are willing to buy. Experimentation, an approach based on taking product assumptions and systematically testing them~\cite{Lindgren2016}, could help these companies on reducing uncertainty, leading to a higher probability of success or, at least, to failure with a smaller consumption of resources. This approach is particularly suitable for software startups given at least two reasons. Fist, Lean Startup, a well-known methodology for startups in the industry based on experimentation~\cite{Frederiksen2017}, claims to be an application of agile software development methods~\cite{Ries2011}. Second, the emergence of cloud computing make the cost of experimenting much smaller than the company had to build the whole infrastructure~\cite{Ewens2018}. Despite these facts, software startups still do not employ experimentation often~\cite{Gutbrod2017}. Research has shown that the use of experimentation in software engineering is hindered by organizational aspects rather than technological limitations~\cite{Lindgren2016}. In software startups, this issue is associated with a high confidence on the idea and a focus on building a perfect product~\cite{Melegati2020}.

To tackle this issue, in a previous study~\cite{Melegati2022}, we developed HyMap, a technique based on cognitive mapping to identify hypotheses on early-stage software startups, as a way to shift the focus of requirements into learning. HyMap consists of a facilitator, following a pre-defined set of questions, interviewing the startup founders in order to create a map depicting the interviewee's understanding of the customers and market. Based on this map, it is possible to generate hypotheses about the product that could be tested to reduce uncertainty. A weakness of the technique is the dependence on an external actor: the facilitator, who might be not available to a software startup interested in using the technique. As previous research has shown, the lack of mentors experienced with innovative companies is another inhibitor to the use of experimentation~\cite{Melegati2019}. 

In this paper, we present \textit{Digital Mentor}, a web-based solution providing HyMap using an automatic conversational bot to replace the human facilitator. We describe the technical solution, the current stage of development, the challenges faced so far, and the next steps.

\section{Background and related work}
\label{sec:related_work}

In this section, we briefly introduce key concepts about conversational bots and how they have been employed in software engineering. Then, we describe HyMap, a technique based on cognitive mapping to identify hypotheses in early-stage startups.

\subsection{Chatbots}

Chatbots, or conversational bots, are software systems that support user interaction via conversation in natural language, normally accessed through the web or social networks~\cite{Perez-Soler2021}. Currently, the available conversational bot solutions rely on NLP-based intent detection, i.e., they try to detect what the user wants by classifying it in abstract classes~\cite{Cabot2021}. These intents can have parameters that define their details called entities. For instance, an intent could be order a pizza and its flavor an entity. These bots are trained with datasets containing examples of intents and entities~\cite{Cabot2021}. These solutions have been adopted in many scenarios in the industry, such as customer support. However, at the current stage, these solutions still converse in a more passive way, responding to the request, rather than following their own initiatives~\cite{Wu2019}. 

\subsection{Chatbots for software engineering practice}

Although, to the best of our knowledge, there are no automatic support to identify hypotheses for software engineering, there have been some proposals or attempts to employ agents in software development. A field in which some proposals have appeared is requirements elicitation. This fact is interesting given the fact that, in our proposal, hypotheses elicitation is just one step of hypotheses engineering, a counterpart of requirements engineering for experiment-driven software development~\cite{Melegati2019he}. Derrick et al.~\cite{Derrick2013} compared the number of requirements gathered in elicitation sessions and how complete they are when groups were using a human facilitator, an automatic agent, or without facilitation. Even though the agent only sputters a pre-defined list of questions without analyzing them, the authors observed better quantity and quality of requirements. Their motivation was also related to ``a lack of access to collaboration professionals such as facilitators and skilled team leaders.'' Rietz and Maedche~\cite{Rietz2019} propose the LadderBot, a chatbot implementing the laddering technique to facilitate requirements elicitation. This process consists of repeatedly asking why questions to the user after initial queries leading the user to think about the requirements. At the reported stage of development, the bot was not able to identify when it should stop asking and relied on the user to ask to stop. The authors had not conducted any evaluation. Another interesting proposal regards a chatbot for agile retrospectives~\cite{Matthies2019}. The bot would take advantage of the artifacts generated during software development to support teams in retrospectives.

\subsection{HyMap}

HyMap is a technique based on cognitive mapping to identify hypotheses about novel software products as those developed by startups~\cite{Melegati2022}. It is based on results showing that startup founders develop an ``implicit theory'' based on their previous experiences to understand and forecast the behavior of customers. Founders rely on this understanding to predict the value and usefulness of a idea for potential customers. Using an adapted cognitive map, a founder, assisted by a facilitator, can depict this ``theory'' in a visual form and use it to generate hypotheses. These hypotheses can be tested in different ways, such as A/B tests, problem, or solutions interviews~\cite{Lindgren2016}. That is, experiments are used here in a broader sense meaning an evidence-based decision making rather than relying on opinions. HyMap consists of a visual language to depict a map representing how a software startup team member, usually the founder, interprets the surroundings, and a defined process to create this map based on a set of questions to be asked to the founder. Table~\ref{tab:hymap_questions} displays the questions and corresponding actions to draw the map.

\begin{table*}
  \caption{HyMap questions}
  \label{tab:hymap_questions}
  \begin{tabular}{cm{0.3\textwidth}m{0.5\textwidth}}
    \toprule
    \#&Question&Action\\
    \midrule
    1 & What is the product name? & Draw an ellipsis at the bottom of the page with the product name. \\
    2 & What are the customers targeted by the solution? & For each customer, draw a circle at the top  with the name. \\
    3 & For each customer: what are the aspects the customer expect to improve using the solution? & For each identified aspect, draw a box below the circles representing customers and connect it to the corresponding customer using an arrow. \\
	4 & Which are the features planned for the solution and which aspect identified in 3 they fulfill? & For each feature, draw a box with dashed lines above the ellipsis representing the product and connect it with arrows to the related aspect and to the product. The arrow connecting the aspect and the feature should be decorated with a label representing the relationship among the concepts. The possible values are ``+'', for a increasing association, ``-'', for a negative one, and ``/o/'' for a neutral. \\
	5 & For each arrow, is there any underlying concept that explains the relationship? & If positive, a new box representing the new concept is added between the two original boxes and it is connected with the two original boxes using the arrows with labels as in 4. This step should be repeated for all arrows, including the new ones, until no boxes are inserted.\\
  \bottomrule
\end{tabular}
\end{table*}

\begin{figure}[h]
	\centering
	\includegraphics[width=\linewidth]{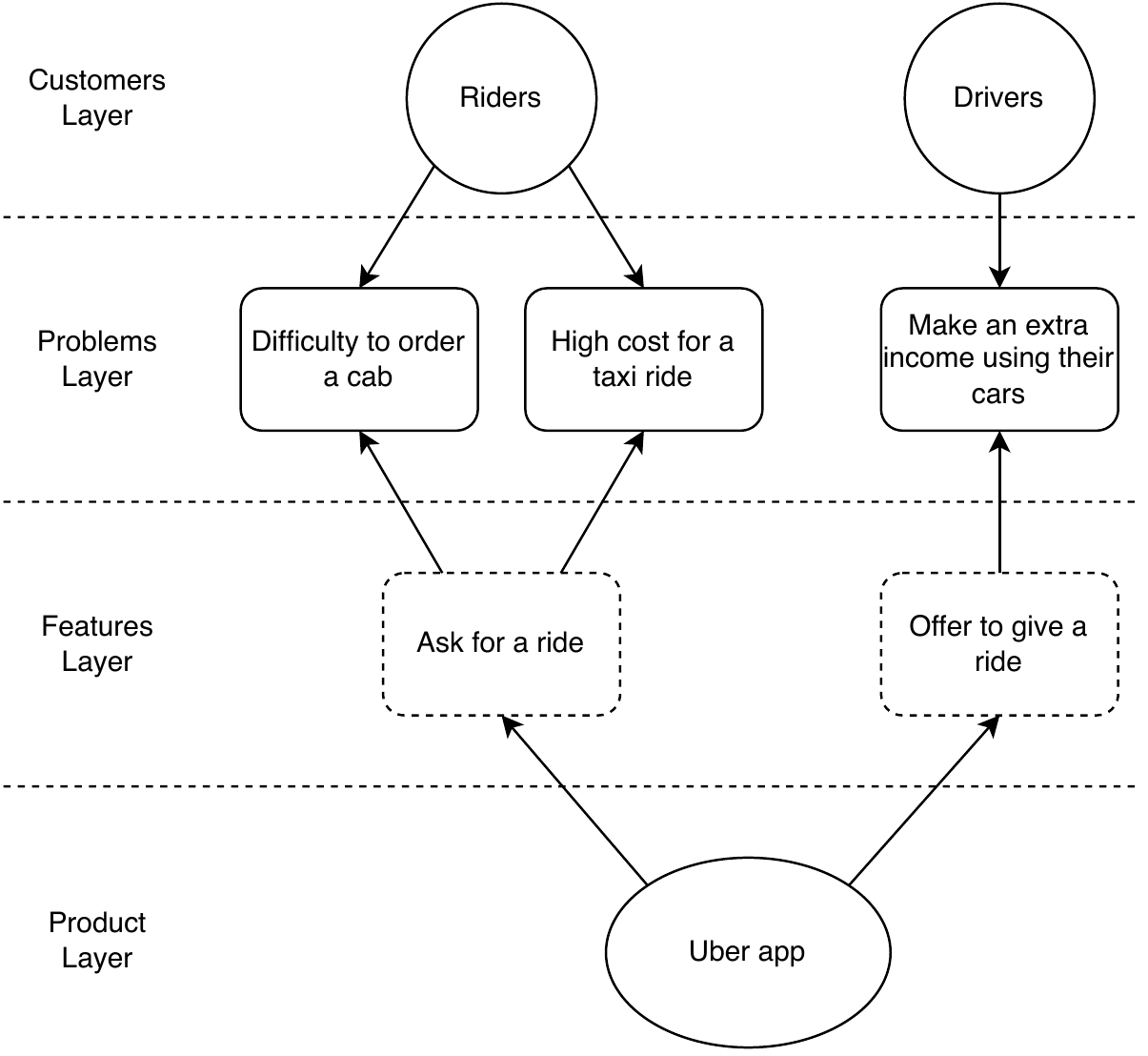}
	\caption{A HyMap for Uber.}
	\label{fig:uber_hymap}
	\Description{A HyMap for Uber presenting the different map layers.}
\end{figure}

As an example, we show a map for the Uber app in Fig.~\ref{fig:uber_hymap}. Of course, since we are already aware that the Uber app is a commercial success, there is no need to draw a map since most of the hypotheses were already validated. However, it is an example easy to understand. We present a simplified version of what would be a HyMap for Uber presenting all the HyMap elements but without covering all product details. The reader could check more complete examples in the paper describing HyMap~\cite{Melegati2022}.

The map is divided into layers where different concepts are presented. In the bottom, the product, in this case, the Uber app, is represented by an oval. The dashed boxes connected to it represent the features in the product. The features are linked to one or more layers of concepts, usually, problems represented by boxes faced by the customers. The customer segments are represented by circles at the top. Each one of the connecting arrows represent a hypothesis that should be tested to check the product viability. For instance, riders have difficulty to order a cab or they face high costs for a ride are to examples. If they were not valid, the product might not be economically viable. 

Arrows connecting different layers represent diverse types of hypotheses and consequently their statements are formed in different forms. Connections between the product and features layers represent feasibility hypotheses, i.e., if the team is able to implement the feature. A template for these hypotheses are: ``the team developing $<$product name$>$ is capable of implementing $<$functionality$>$.'' Arrows between feature and problem layers, or between different problem layers, lead to value hypotheses. In this case, a template is: ``$<$Functionality or problem$>$ $<$increases, decreases or does not affect$>$ $<$problem$>$.'' Finally, arrows between customer and problem layers represent problem hypotheses, i.e., if the represented issues are real problems for the customers. A template is: ``$<$Customer segment$>$ $<$has/would like to$>$ $<$problem$>$.''

The dependence of HyMap on a human facilitator to ask the questions is a weakness. The lack of mentors experienced with startups is a known inhibitor to the adoption of experimentation for software startups~\cite{Melegati2019}. Replacing the human facilitator by a digital one has several advantages First, it simplifies the access to the technique. The team does not have to find a facilitator that might not be available, especially, in less developed ecosystems. Second, it allows teams to increase the frequency of running HyMap without depending on the facilitator availability. For instance, if the team decides to change the idea, doing a ``pivot'' in startup jargon, they can do another session. Second, 

\section{Proposed solution and current state of development}
\label{sec:current_state}

\begin{figure}[h]
  \centering
  \includegraphics[width=\linewidth]{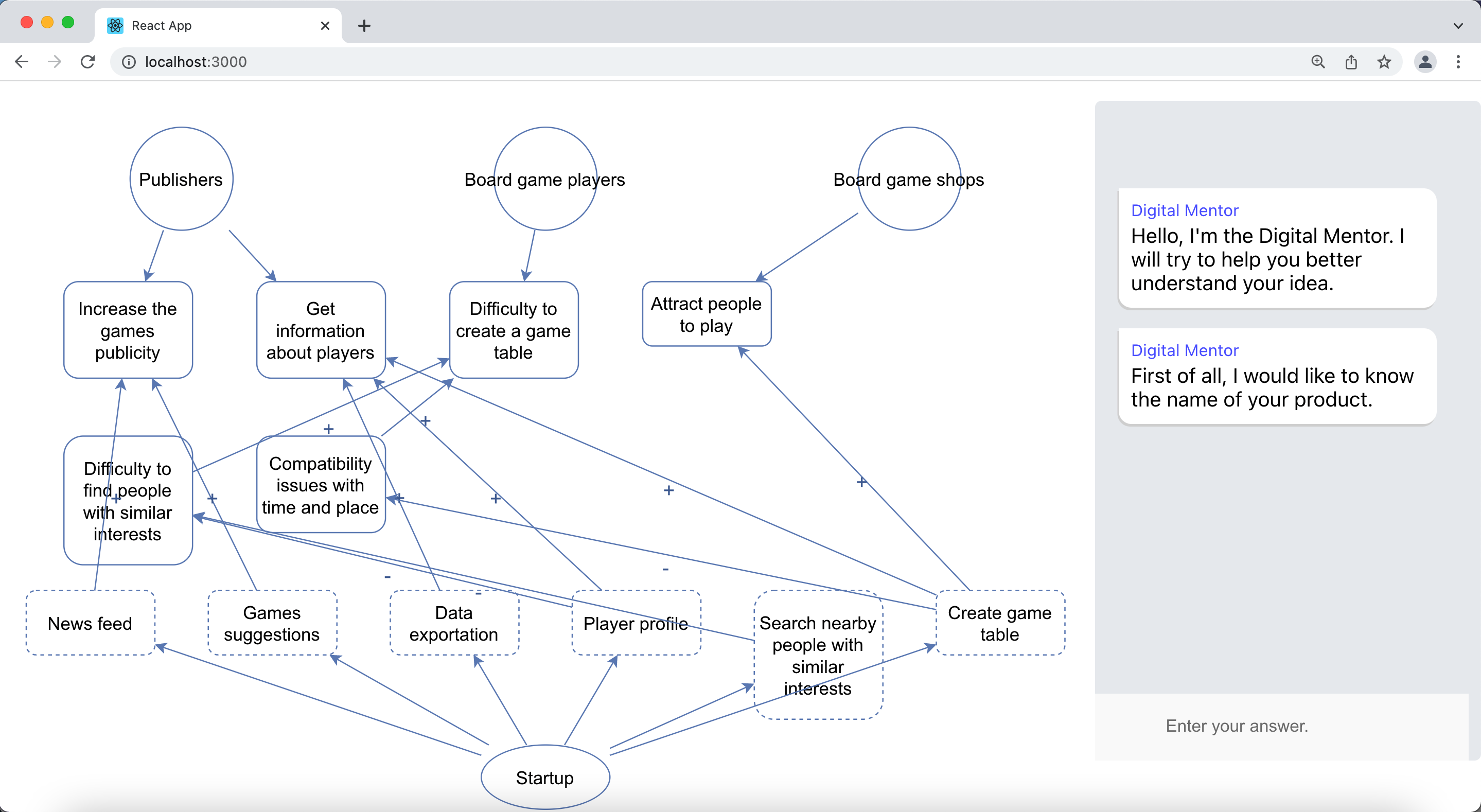}
  \caption{Digital Mentor.}
   \label{fig:webpage}
  \Description{A webpage showing a graph and a chat window.}
\end{figure}

\textit{Digital Mentor} is a single-page web-based system developed using ReactJS. The backend consists of a Rasa\footnote{https://rasa.com/} Natural Language Understanding (NLU) service. The webpage, as shown in Fig.~\ref{fig:webpage}, consists of a chat window on the right where the users interacts with the bot and an area on the left where the cognitive map is drawn. Given that the bot is responsible for drawing the map, the conversation flow is implemented in the front end using a state machine. Each input from the user is sent to backend NLU that determines its type. Based on the return, the frontend fills an internal representation of the map, updates it on the screen, and changes to the next state. Some special return types, such as asking for clarifications, trigger specific utterances from the bot.

The bot's goal is to extract from the user's answers pieces of text to fulfill the map elements in a fluent way enabling the generation of coherent hypothesis statements. It should be able to handle different ways the user might answer the questions. For instance, in the step to identify the reasons customers would be interested in the product, the bot asks ``Why would riders want to use Uber?" and the user can use different structures to answer: from more direct ones, e.g., ``to book a ride'', to more elaborated ones, e.g., ``because it is hard to find a cab in some places.'' Given the needs for this bot, we decided to use Rasa's DIET classifier including not only the word representation but also the part of speech (POS). In such a way, it would be possible to differentiate between concepts represented by noun clauses, e.g., ``it was difficult to find a cab'', and verb clauses, e.g., ``they want to take a ride.'' This difference is essential to select the suitable template for the hypothesis statements. 

The state machine implements the questions in Table~\ref{tab:hymap_questions} in an adapted fashion to expect simpler answers. For instance, for Question 2, regarding the customer types, the bot asks one at a time, followed by a question if the user wants to add another customer or not. This process continues until the user answers that no more customers is needed. Similarly, when asking about the problems and relationships, the bot asks in a second moment the nature of the relationship (positive, negative or neutral).

The NLU is trained to identify when the user needs help to understand a question. In this case, for each state, we defined an utterance to better detail what is expected as an answer for that question. That is, this intention occurs similar to an interruption. For each state aiming to obtain information to fill the map, i.e., excluding those to control the flow, we trained the NLU to detect the intent. For the states to identify problems, we defined two different intents: desire description and difficulty description. They differ on how the respective hypotheses will be built. In the hypotheses statement, the desire is generally preceded by the verb \textit{to want}. Meanwhile, for a difficulty, the clause starts with the word \textit{difficulty}.

\section{Challenges and next steps}
\label{sec:challenges}

In the current state of development, the NLU is not able to always correctly identify the intent as designed. However, in most cases, given the state machine, the frontend is able to use the result in a way to fulfill the map elements even though, in some situations, the hypotheses is not coherently formed. We recognized some issues hindering better results that we describe below.

First, there are answers syntactically similar but associated with different intents depending on semantics. For instance, consider two sentences used to train two intents. The first is ``They want to [attract people to play]'', it should be classified as a desire intent and the piece in brackets should be extracted as a desire entity. The second is ``The app allows the users to [export data],'' it should be considered a feature description intent and the piece in brackets the feature entity. From a syntactic point of view, the sentences are similar, especially when considering the possible values for the entities and structures employed by the user. In this case, it is hard to train the NLU to distinguish between these answer types. Specially given the second issue, the broadness of possible topics. Since our goal with the bot is to help any software startup, there is a wide range of possible subjects on which the bot should be trained. For instance, a conversational bot to get orders from a restaurant will be trained with the vocabulary needed for that restaurant and the ordering process. This breadth of topics also hinders the possibility of employing a pre-defined list of values to specific entities.  

Finally, the problems discussed before are more evident given the lack of a dataset of possible answers. This issue is associated with the novelty of systematically handling hypotheses for experimentation and the consequent lack of  startups that have used hypotheses to be the source of training data. 

To overcome these issues, our current plan is to evaluate the possibility of using a smaller number of intents and relying more on the state machine. In this case, the NLU might only be responsible for identifying interruptions and clause types to create coherent hypothesis statements. For the long term, this bot is just the first step towards our vision of developing a digital mentor including several attributes of human startups mentors, such as open-mindedness and motivational. 

\section{Conclusions}
\label{sec:conclusions}

To improve their chance of success and reduce the waste in case of failure, software startups should identify the hypotheses on which their ideas are based and test them using diverse types of experiments. HyMap is a technique based on cognitive mapping in which a facilitator helps founders to depict their understanding into a graphical form and extract hypotheses from it. We propose to replace the human facilitator with a conversational bot that asks the pre-defined questions of HyMap, draws the respective cognitive map, and builds the corresponding hypothesis statements. At the current stage of development, the NLU in which the chatbot is based struggles in identifying the correct intent but the state machine generally adjusts the chat to the proper flow. We discussed how we plan to move forward and reach a reliable solution. 

\bibliographystyle{ACM-Reference-Format}
\bibliography{refs}

\end{document}